\documentclass[aip,%
 jmp,%
 amsmath,amssymb,
 reprint,
 numerical%
]{revtex4-1}

\usepackage{graphicx}
\usepackage{dcolumn}
\usepackage{bm}

\begin{document}

\title{Influence of Carrier-Carrier Scattering on Electron Transport in Monolayer Graphene}

\author{X. Li}
\affiliation{Department of Electrical and Computer Engineering, North Carolina State University, Raleigh, NC 27695-7911}

\author{E. A. Barry}
\affiliation{Department of Electrical and Computer Engineering, North Carolina State University, Raleigh, NC 27695-7911}

\author{J. M. Zavada}
\affiliation{Department of Electrical and Computer Engineering,
North Carolina State University, Raleigh, NC 27695-7911}

\author{M. Buongiorno Nardelli}
\affiliation{Department of Physics, North Carolina State University,
Raleigh, NC 27695-8202}
\affiliation{CSMD, Oak Ridge National
Laboratory, Oak Ridge, TN 37831}

\author{K. W. Kim}
\affiliation{Department of Electrical and Computer Engineering,
North Carolina State University, Raleigh, NC 27695-7911}

\begin{abstract}
The influence of electron-electron scattering on the distribution function and transport characteristics of intrinsic monolayer graphene is investigated via an ensemble Monte Carlo simulation. Due to the linear dispersion relation in the vicinity of the Dirac points, it is found that pair-wise collisions in graphene do not conserve the ensemble average velocity in contrast to conventional semiconductors with parabolic energy bands.  Numerical results indicate that electron-electron scattering can lead to a decrease in the low field mobility by more than 80~\% for moderate electron densities. At high densities, the impact gradually diminishes due to increased degeneracy.

\end{abstract}

\maketitle

Since graphene was first realized experimentally in
2004,~\cite{Novoselov2004} it has attracted significant
interest due to its unique electronic properties. At low electron
energies near the inequivalent Dirac points, there is no gap
between the valence and conduction bands and the dispersion of the
energy bands is linear.~\cite{CastroNeto2009} Extremely high
electrical mobilities have been reported in suspended graphene,
exceeding 10$^5$ cm$^2$/Vs near room
temperature,~\cite{Bolotin2008351} suggesting potential applications to ultrahigh frequency electronic devices.~\cite{Y.-M.Lin02052010} The carrier density, which is controlled by the gate voltage, can be expected to vary by
orders of magnitude.  An interesting consequence of the linear energy dispersion is that the ensemble average velocity is not necessarily conserved upon an electron-electron (e-e) scattering event.  Accordingly, inter-carrier collisions deserve careful consideration in the determination of the transport properties.

Das Sarma {\em et al.}~\cite{DasSarma2007_Hwang2007} found the
inelastic e-e scattering rate and mean free path in graphene
through the analysis of the quasiparticle self-energies. The
scattering rate, calculated for electron densities from $1 - 10
\times 10^{12}$~cm$^{-2}$, was found to be of the same order of
magnitude for electron-phonon scattering rates evaluated in the
deformation potential approximation (DPA).~\cite{Shishir2009}
Several authors have considered the electronic transport properties of graphene based on approaches such as the Monte Carlo simulation (see, for example, Ref.~\onlinecite{Goldsman2008});  however, the effects of e-e scattering have yet to be addressed to the best of our knowledge.  In the present study, we  examine the influence of this interaction mechanism in intrinsic monolayer graphene at room temperature.  A full-band ensemble Monte Carlo method is used for accurate analysis of the distribution function as well as its macroscopic manifestations, particularly, the electron low-field mobilities and drift velocities.

In both bulk and two-dimensional (2D) conventional semiconducting systems, e-e scattering has been well studied as is documented in the literature.~\cite{Bonno2005_Fischetti1988_Goodnick1988}
During an e-e scattering event, both the energy and momentum are conserved.
In a parabolic band structure, common in conventional
semiconductors, momentum conservation directly leads to the
conservation of velocity. This can be readily shown by multiplying the
momentum conservation equation by ${\hbar}/{m}$, which gives
$\mathbf{v_{1}}+\mathbf{v_{2}}=\mathbf{v^{\prime}_{1}}+\mathbf{v^{\prime}_{2}}$.
It is then clear that, in a material with a parabolic band
structure, e-e scattering has no direct effect on the drift
velocity (which is an ensemble averaged quantity). The situation becomes different in graphene, where the
energy dispersion in the region of the inequivalent Dirac points
is linear, therefore removing the constraint that momentum
conservation leads directly to velocity conservation.

The problem of e-e scattering in nonparabolic bands was previously
treated by Bonno {\em et al.}~\cite{Bonno2008} Here we use a
similar approach, making it specific to the linear band structure
of graphene. The transition probability for two electrons, with
wavevectors $\mathbf{k_{1}}$ and $\mathbf{k_2}$, is given by
Fermi's golden rule
\begin{eqnarray}
\label{See}
s(\mathbf{k_{1}},\mathbf{k_{2}};\mathbf{k_{1}^\prime},\mathbf{k_{2}^\prime}) = &&\frac{2\pi}{\hbar}|M|^{2}\delta[E(\mathbf{k^\prime_{1}}) + E(\mathbf{k^\prime_{2}})\nonumber\\ 
&&-E(\mathbf{k_{1}})-E(\mathbf{k_{2}})]\,,
\end{eqnarray}
where $E(\mathbf{k})$ is the electron energy dispersion relation and $\mathbf{k^\prime_{1}}$ and $\mathbf{k^\prime_{2}}$ are the wavevectors of the final states after scattering.  The matrix element of interaction $|M|$, when accounting for exchange scattering (i.e., the indistinguishability of electron pairs with like spin after collision), is given by~\cite{Moskova}
\begin{equation}
|M|^{2} = \frac{1}{2}[|V(q)|^{2} + |V(q^\prime)|^{2} -V(q)V(q^\prime)]\,,
\end{equation}
where the Coulomb scattering matrix $V(q)$ between two electrons ($\mathbf{k_{1}} \rightarrow \mathbf{k^\prime_{1}}$; $\mathbf{k_{2}} \rightarrow \mathbf{k^\prime_{2}}$) is
\begin{equation}
V(q)=\frac{2\pi
e^{2}}{\epsilon(q)qA}\frac{1+\cos(\theta_{\mathbf{k_{1}}\mathbf{k^\prime_{1}}})}{2}\frac{1+\cos(\theta_{\mathbf{k_{2}}\mathbf{k^\prime_{2}}})}{2} \,,
\end{equation}
with $q=|\mathbf{k_{1}}- \mathbf{k^\prime_{1}}|$, $\epsilon(q)$ is the static dielectric function, $\theta_{\mathbf{k}\mathbf{k^\prime}}$ is the angle
between the wavevectors $\mathbf{k}$ and $\mathbf{k^\prime}$,  and $A$ is the normalization area.  The corresponding element $V(q^\prime)$ ($\mathbf{k_{1}} \rightarrow \mathbf{k^\prime_{2}}$; $\mathbf{k_{2}} \rightarrow \mathbf{k^\prime_{1}}$) can be expressed likewise with $ q^\prime=|\mathbf{k_{1}}- \mathbf{k^\prime_{2}}|$. The dielectric function, which is valid for densities larger than $n=1\times 10^{12}$~cm$^{-2}$, is considered in the random-phase approximation.~\cite{Hwang2007}
\begin{figure}
\includegraphics[]{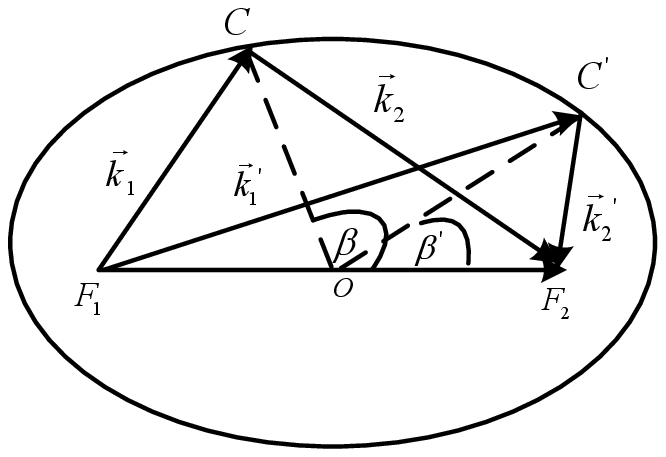}
\caption{Schematic representation of the
initial ($\mathbf{k_{1}},\mathbf{k_{2}}$) and final
($\mathbf{k_{1}^\prime},\mathbf{k_{2}^\prime}$) electron pair
states which satisfy energy and momentum conservation in the
linear band structure of graphene.  $\beta$ is defined as the
angle between the long axis and the line $\overline{OC}$ and
$\beta^\prime$ between the long axis and the line
$\overline{OC^\prime}$.}\label{ee-scat-fig1}
\end{figure}

The scattering rate, including the effect of degeneracy, is then
obtained as the double sum over all the final states
$\mathbf{k^\prime_{1}}$ and the partner electrons $\mathbf{k_{2}}$ (with $\mathbf{k^\prime_{2}}$ determined by momentum conservation)
\begin{eqnarray}
\label{tauee}
\frac{1}{\tau_{ee}(\mathbf{k_{1}})}=&&\sum_{\mathbf{k_{2}}}\sum_{\mathbf{k^\prime_{1}}} s(\mathbf{k_{1}},\mathbf{k_{2}};\mathbf{k_{1}^\prime},\mathbf{k_{2}^\prime})
f(\mathbf{k_{2}})\nonumber\\
&&\times
[1-f(\mathbf{k^\prime_{1}})][1-f(\mathbf{k^\prime_{2}})]\,,
\end{eqnarray}
where the $f$'s are the occupation probabilities. Within the Monte
Carlo simulation, the Pauli exclusion principle is treated using
the standard rejection technique.

In graphene, the energy dispersion of the conduction band is
linear in the vicinity of the Dirac points; i.e., $E=\hbar v_{f}|\mathbf{k}|$, with the Fermi velocity $v_{f}$ ($\simeq 10^8$~cm/s). Therefore, the
electrons behave as massless particles and can not be analyzed by
the center of mass concept.~\cite{Sabio2010}  However, by
considering the equations of momentum and energy conservation, it
can be shown that all the possible final states after scattering
lie on an ellipse defined by $\mathbf{k_{1}}+\mathbf{k_{2}} = \mathbf{k^\prime_{1}}+\mathbf{k^\prime_{2}}$ and
$|\mathbf{k_{1}}| + |\mathbf{k_{2}}| = |\mathbf{k^\prime_{1}}| + |\mathbf{k^\prime_{2}}|$.  As shown in Fig.~\ref{ee-scat-fig1}, the initial point (i.e., the coordinate centrum at the $K$ or $K^\prime$ point in the Brillouin zone) and the terminal point of the vector $\mathbf{k_{1}}+\mathbf{k_{2}}$ specify the two foci of the ellipse, with the length of long axis of the ellipse determined by
$|\mathbf{k_{1}}|+|\mathbf{k_{2}}|$.
\begin{figure}
\includegraphics[]{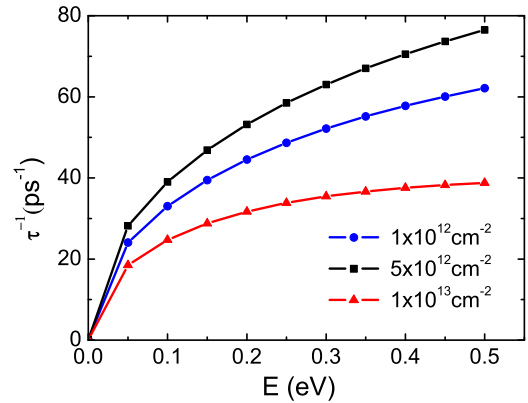}
\caption{(Color online) Electron-electron scattering rate for electron concentrations $n$ = 1$\times$10$^{12}$~cm$^{-2}$ (circle), $n$ = 5$\times$10$^{12}$~cm$^{-2}$ (square), and
$n$ = 1$\times$10$^{13}$~cm$^{-2}$(triangle).}\label{ee-scat-fig2}
\end{figure}

Our Monte Carlo model applied in the simulation includes the full electronic band structure as calculated in the tight binding approximation.  The
electron-phonon scattering rates and phonon dispersion relations,
previously obtained from a first principles approach based on
density functional perturbation theory, are included for all six
branches.~\cite{Borysenko2010} All electron-phonon transition
possibilities, including intravalley and intervalley, are
accounted for as well as the e-e scattering. Finally, since our simulations include large electron densities we account for degeneracy, where Pauli
exclusion is non-negligible, by implementation of the rejection
technique in the selection of the final state after
scattering.~\cite{Lugli1985}
\begin{figure}
\includegraphics[]{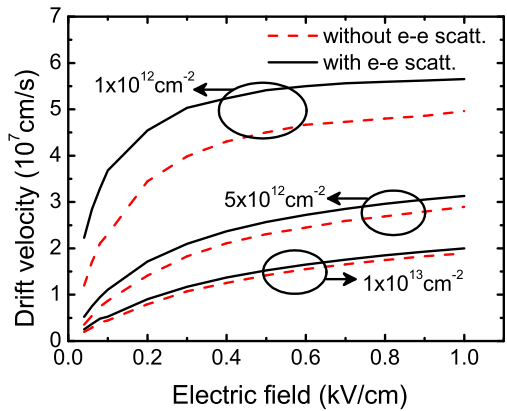}
\caption{(Color online) Drift velocity
for $n$ = 1$\times$10$^{12}$~cm$^{-2}$, $n$ = 5$\times$10$^{12}$~cm$^{-2}$, and $n$ =
1$\times$10$^{13}$~cm$^{-2}$ with (dashed line) and without (solid
line) e-e scattering.}\label{ee-scat-fig3}
\end{figure}

In Fig.~\ref{ee-scat-fig2}, we show the calculated e-e scattering rate, assuming a 
Fermi-Dirac distribution of initial states and that all final states are all available.
The scattering rate is found to have a maximum of approximately 40-80~ps$^{-1}$, for the energy range
and electron densities considered, which is at least an order of magnitude larger than the
total electron-phonon scattering rate.~\cite{Borysenko2010}
In Fig.~\ref{ee-scat-fig3}, we examine the effect
of e-e scattering on the drift velocity from the low-field to saturation regimes. It is evident that there is a more than 80~$\%$ degradation of the low field mobility, which can be estimated by taking the slope of the drift velocity in the low field region, for an electron concentration
of $n=1\times 10^{12}$~cm$^{-2}$. Additionally, there is a degradation of the saturation drift velocity by
$\sim$16~$\%$. For low fields, when electrons rarely gain enough
energy to emit an optical phonon, acoustic phonon and e-e scattering events dominate the electron dynamics and determine the mobility.~\cite{DasSarma2007_Hwang2007} As the field is raised, optical phonon scattering becomes more prevalent and the effect of e-e scattering is somewhat mitigated. The dominance of e-e scattering at low energies, relative to the electron-phonon scattering rate, suggests that other scattering mechanisms (i.e.,  substrate dependent surface polar phonons, remote impurities, etc.) may tend to wash out the resulting degradation.
\begin{figure}
\includegraphics[]{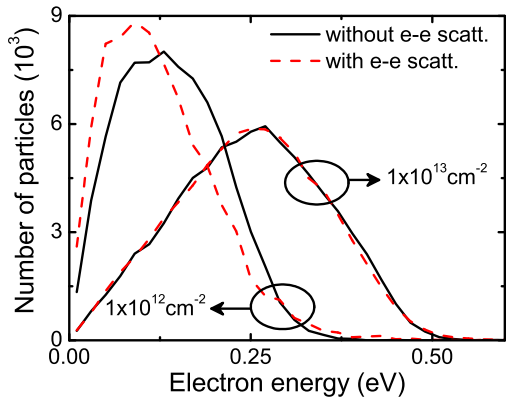}
\caption{(Color online) Energy
distribution function for $n$ = 1$\times$10$^{12}$~cm$^{-2}$ and
$n$ = 1$\times$10$^{13}$~cm$^{-2}$ with (dashed line) and without
(solid line) e-e scattering. \label{ee-scat-fig4}}
\end{figure}

For larger electron densities, $n=5\times 10^{12}$~cm$^{-2}$ and $n=1\times 10^{13}$~cm$^{-2}$, this effect is suppressed by the reduction of available states with high and low energies. This can be seen from the electron energy distribution functions, given in Fig.~\ref{ee-scat-fig4}, at an electric field of $E=1$ kV/cm for the highest and lowest carrier densities considered. Just as in conventional semiconductors, e-e scattering results in an extension of the thermal tail of the distribution function, while simultaneously shifting its peak backward. Consequently, though an increase in electron density results in an increased e-e scattering rate, the
distribution function is less affected than in the moderate density
case. This results in a larger degradation of the low-field mobility and saturation drift velocity for lower concentrations, as is evident in Fig.~\ref{ee-scat-fig3}.
\begin{figure}
\includegraphics[]{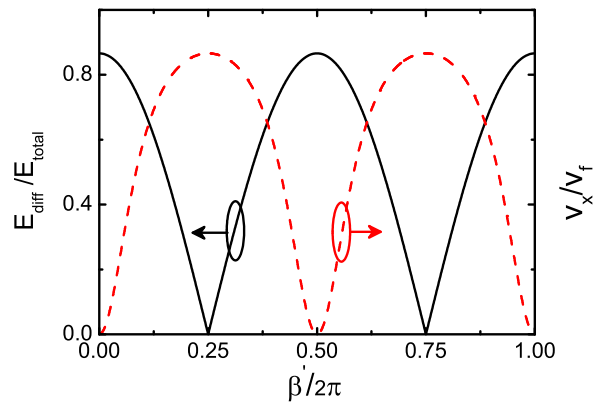}
\caption{(Color online) Calculated
$E_{\rm diff}/E_{\rm total}$ (solid line) and $v_x/v_f$ (dashed line)
assuming the long axis of the ellipse is along the direction of
the electric field.}\label{ee-scat-fig5}
\end{figure}

Figure~\ref{ee-scat-fig5} presents how the selection of final
state pairs effects the velocity of the ensemble which, as stated previously, is not necessarily conserved. On average, the momentum of an initial electron pair will be directed along the electric field which corresponds to the long axis of the ellipse, in Fig.~\ref{ee-scat-fig1}. The curve corresponding to
$E_{\rm diff}/E_{\rm total}$, where $E_{\rm diff}=|E_1^\prime-E_2^\prime|$ is
the energy difference of the final state electron pair, and
$E_{\rm total}=E_1^\prime+E_2^\prime$, has maximal values when
$\beta^\prime$ is equal to $0$ or $\pi$. Also shown in
Fig.~\ref{ee-scat-fig5}, is the value of the final state drift
velocity in the $x$ direction, $v_x=v_{1x}^\prime+v_{2x}^\prime$, 
normalized by the fermi velocity $v_f$. Opposite to $E_{\rm diff}/E_{\rm total}$, the velocity
in the direction of the electric field has minima at
$\beta^\prime$ is equal to $0$ or $\pi$. This results in a
decrease of the the ensemble drift velocity for moderate electron
densities, on the order of $n$ = 1$\times$10$^{12}$~cm$^{-2}$, since there are available high and low energy states to scatter into. At larger densities, the lowest energy states are already mostly
occupied, thereby preventing transition to the energy states which most greatly reduce
the drift velocity. As a result, the effect of e-e scattering is most pronounced for $n$ = 1$\times$10$^{12}$~cm$^{-2}$, and largely suppressed for $n$ = 1$\times$10$^{13}$~cm$^{-2}$. 

In summary, our results clearly indicate that due to e-e scattering, there is an electron density dependent degradation of the low-field mobility and saturation drift velocity in intrinsic monolayer graphene. This effect is most evident at moderate electron densities, while being suppressed at higher densities due to the unavailability of low energy states. For moderate electron densities, it is predicted that the low-field mobility and saturation drift velocity are degraded on the order of 80~\% and 16~\%, respectively.
   
This work was supported, in part, by the DARPA/HRL CERA, US ARO, and SRC/FCRP
FENA programs.

\end{document}